\documentclass[sigconf]{acmart}
 
\acmConference[EASE 2025]{EASE 2025}{June 17-20, 2025}{Istanbul, Turkey}
\AtBeginDocument{%
  }

\acmISBN{978-1-4503-XXXX-X/2018/06}

\usepackage{amsmath,amsfonts}
\usepackage{algorithmic}
\usepackage{graphicx}
\usepackage{textcomp}
\usepackage{xcolor}
\usepackage{booktabs}
\usepackage{natbib}
\usepackage[most]{tcolorbox}
\usepackage[explicit]{titlesec}

\copyrightyear{2025}
\acmYear{2025}
\setcopyright{rightsretained}
\acmConference[EASE '25]{Proceedings of the 2025 Evaluation and Assessment in Software Engineering}{June 17--20, 2025}{Istanbul, Turkiye}
\acmBooktitle{Proceedings of the 2025 Evaluation and Assessment in Software Engineering (EASE '25), June 17--20, 2025, Istanbul, Turkiye}
\acmPrice{}
\acmDOI{10.1145/3756681.3757030}
\acmISBN{979-8-4007-1385-9/25/06}

\begin{document}

\title{Toward Inclusive Low-Code Development: Detecting Accessibility Issues in User Reviews}


\author{Mohammadali Mohammadkhani}
\affiliation{%
  \institution{Sharif University of Technology}
  \city{Tehran}
  \country{Iran}}
\email{mo.mohammadkhani@sharif.edu}

\author{Sara Zahedi Movahed}
\affiliation{%
  \institution{Sharif University of Technology}
  \city{Tehran}
  \country{Iran}
}
\email{sara.zahedi@sharif.edu}

\author{Hourieh Khalajzadeh}
\affiliation{%
 \institution{Deakin University}
 \city{Melbourne}
 \country{Australia}}
\email{hkhalajzadeh@deakin.edu.au}

\author{Mojtaba Shahin}
\affiliation{%
  \institution{RMIT University}
  \city{Melbourne}
  \country{Australia}}
\email{mojtaba.shahin@rmit.edu.au}

\author{Khuong Tran Hoang}
\affiliation{%
 \institution{Deakin University}
 \city{Melbourne}
 \country{Australia}}
\email{Khuong.Tran@alumni.uts.edu.au}

\renewcommand{\shortauthors}{Mohammadkhani et al.}

\begin{abstract}
Low-code applications are gaining popularity across various fields, enabling non-developers to participate in the software development process. However, due to the strong reliance on graphical user interfaces, they may unintentionally exclude users with visual impairments, such as color blindness and low vision. This paper investigates the accessibility issues users report when using low-code applications. We construct a comprehensive dataset of low-code application reviews, consisting of accessibility-related reviews and non-accessibility-related reviews. We then design and implement a complex model to identify whether a review contains an accessibility-related issue, combining two state-of-the-art Transformers-based models and a traditional keyword-based system. Our proposed hybrid model achieves an accuracy and F1-score of 78\% in detecting accessibility-related issues.
\end{abstract}




\maketitle

% \titlespacing{\section}{0pt}{\parskip}{-\parskip}
% \titlespacing{\subsection}{0pt}{\parskip}{-\parskip}
% \titlespacing{\subsubsection}{0pt}{\parskip}{-\parskip}

\titlespacing\section{0pt}{6pt plus 2pt minus 2pt}{0pt plus 2pt minus 2pt}
\titlespacing\subsection{0pt}{6pt plus 2pt minus 2pt}{0pt plus 2pt minus 2pt}
% \titlespacing\subsubsection{0pt}{6pt plus 2pt minus 2pt}{0pt plus 2pt minus 2pt}

\section{Introduction}
The International Organization for Standardization (ISO) defines accessibility in interactive systems as \textit{“the usability of a product, service, environment, or facility by people with the widest range of capabilities”} \cite{international2008ergonomics}. A similar definition applies to web applications \cite{petrie2015towards}. These definitions emphasize that every user should have an equal, non-discriminatory experience when using systems. Ensuring software accessibility and inclusiveness is vital to maximize the usability of applications for a broad user base. However, software engineering research has mainly focused on the accessibility of software products, with less attention to engineering processes and development tools \cite{el2023evaluating}.

Low-code applications are an important subcategory of software applications, classified as development tools. Low-code applications support rapid application development, one-step deployment, execution and management \cite{vincent2019magic}. They use declarative, high-level programming abstractions, such as model-driven and metadata-based programming languages \cite{vincent2019magic}. Low-code platforms typically provide a graphical interface where users can drag and drop components, automate workflows, and configure settings to design applications instead of writing extensive manual code from scratch. Low-code platforms relying heavily on graphical user interfaces can often exclude users with visual impairments, including color blindness and low vision \cite{KHALAJZADEH2025107570}.

Users may share their concerns regarding the accessibility and inclusiveness of low-code applications through app reviews. However, low-code developers must read through all app reviews to identify which ones discuss accessibility issues. Manually detecting these issues can be tedious, time-consuming, and prone to errors. An accurate, automated solution can assist developers in identifying and resolving these issues more efficiently. To fill this gap, in this paper, we analyze the accessibility-related issues in low-code application reviews and propose a way to automatically detect such issues. Therefore, we formulated the following research question.
\begin{center}
\label{sec:RQ}
\begin{tcolorbox}[arc=1mm,width=1.0\columnwidth,
                  top=1mm,left=1mm,  right=1mm, bottom=1mm,
                  boxrule=.75pt]%[colback=white!2!white,colframe=black!75!black]
{\textbf{RQ}: How can we accurately and automatically detect accessibility-related issues in low-code application reviews?}
\end{tcolorbox}
\end{center}

To address this research question, we first compile a dataset by combining two methods: developing a crawler module to gather data from a low-code app review website\footnote{www.trustradius.com} and incorporating a subset of reviews from pre-existing datasets, including a dataset from accessibility user reviews \cite{aljedaani2021learning} and a dataset of the reviews collected from G2 website\footnote{https://www.g2.com/}. Then, we manually annotate these collected app reviews based on the criteria of having accessibility-related issues and concerns. After the data collection phase, we design and implement a complex model that leverages the Transformers architecture and neural networks, combined with a traditional keyword-based system. Our model takes an app review as input and predicts its label based on whether it contains accessibility-related issues. Finally, we evaluate our hybrid model, along with three other state-of-the-art models, on test data, and the results show that our model outperforms the others across multiple evaluation metrics. Our contribution consists of three main parts:
\begin{itemize}
    \item Gathering and annotating a dataset of 4,762 low-code app reviews, and releasing it publicly \cite{anonymous_2024_14060778}.
    \item Designing and implementing a complex model to detect accessibility-related issues in low-code app reviews.
    \item Implementing a crawler module to collect app reviews for dataset construction.
\end{itemize}

\section{Proposed Model}
\label{sec:proposedmethod}
In this section, we discuss the details of our methodology to answer our research question \ref{sec:RQ}.

\subsection{A Dataset of Accessibility-related Reviews}
To develop our model, we needed a labeled dataset of low-code app reviews. In this section, we go through the details of our dataset construction.

\subsubsection{Collecting Low-Code App Reviews: }
In the first step, we needed to collect low-code app reviews. Three steps were used to gather a proper collection of low-code app reviews. 

\textbf{Step 1.} We initially used a dataset from \cite{aljedaani2021learning}, containing 2,664 accessibility-related software app reviews. However, since it includes reviews of general apps rather than exclusively low-code platforms, we did not adopt it as the primary source. Instead, we randomly selected 400 reviews from this source to expose our model to a broader range of accessibility-related cases, improving its ability to identify both general and low-code-specific accessibility issues. To ensure data quality, we conducted a manual review, filtering out redundant, excessively brief, or low-content reviews and replacing them with more informative alternatives from the same source. Ultimately, we managed to incorporate 398 reviews from this source into our final dataset.

\textbf{Step 2.} To center our dataset around low-code app reviews, we collected low-code application reviews from \textbf{trustradius}\footnote{https://www.trustradius.com} website, which provides a comprehensive set of evaluations for low-code applications. The website's robots.txt file explicitly permitted us to crawl its reviews page\footnote{https://www.trustradius.com/robots.txt} and the crawling process adhered to the guidelines outlined in this file. 
We used the \textbf{BeautifulSoup} module in \textbf{bs4}\footnote{https://pypi.org/project/beautifulsoup4/} Python library for the crawling process. Using the regex and functions provided by BeautifulSoup, we identified and saved the links to the review pages. Subsequently, we stored the reviews from each page into a text file within our local project, which we later used in our final dataset. It is crucial to note that the reviews were formatted as bullet points, listing the pros and cons of the low-code applications. Consequently, most of the reviews were not complete sentences and often consisted of phrases indicating the application's positive or negative aspects. This step led to the collection of 4,200 low-code app reviews. The code we developed for the crawling process is published publicly \cite{anonymous_2024_14060778}.

\textbf{Step 3.} In this final step, we aimed to enrich our dataset and extract further low-code platform reviews. To do so, we first identified the top 10 low-code platforms based on recommendations from \cite{robmarvinlowcode}, as Appian, Mendix, Zoho Creator, OutSystems, Google App Maker, QuickBase, TrackVia, Salesforce Platform, Microsoft PowerApps and Nintex Workflow. Using the \textbf{Web Automation} tool\footnote{https://webautomation.io/} and scripts built in Python with regex, we then extracted data from commercial peer-to-peer review websites like \textbf{G2}\footnote{https://www.g2.com/} and \textbf{Gartner}\footnote{https://www.gartner.com/peer-insights/home} and filtered out relevant users'reviews on those selected low-code platforms. G2 and Gertner employ a strict and precise review moderation process to ensure the authenticity and relevance of published reviews, minimizing spam and low-quality submissions. From the dataset extracted from these sources, we sampled 20\% of the app reviews for each low-code application. We then refined this selection by the same process applied in Step 1 and after this curation step, we incorporated a total of 164 app reviews from this source into our final dataset.

Ultimately, we collected and preprocessed (See \ref{sec:preprocessing} for more details on data cleaning and preprocessing) 398 reviews from Step 1, 4200 from Step 2, and 164 low-code app reviews from Step 3.

\subsubsection{Constructing Accessibility-related Low-Code Reviews Dataset: }
\label{sec:dataset}
Our next task was to categorize the 4,762 collected low-code app reviews as either accessibility-related or non-accessibility-related. To enhance the reliability of the labeling process, we established a set of guidelines before beginning the annotation. Although the reviews collected in Step 1 had already been labeled accessibility-related reviews, we relabeled them according to these guidelines to ensure consistency with the reviews collected in other steps:

\underline{Rule 1}: Our primary goal is to detect accessibility-related issues in reviews. Therefore, reviews indicating that everything is fine with the app and reporting no issues or bugs should be labeled as 0, signifying the absence of accessibility-related problems.

\underline{Rule 2}: Reviews that report a bug or inconvenience that affects users and renders the app or some of its features inaccessible are labeled 1, indicating the presence of accessibility issues. Examples in our dataset include problems with the user interface, navigation, customization, usability, and similar aspects.

\underline{Rule 3}: Reviews mentioning problems related to the developer side, rather than the client side, are labeled 0 because our goal was to detect accessibility issues from the users of low-code apps. Examples of such issues include connection problems with APIs and databases, integration challenges within the codebase, and automation issues in the development and testing processes.

The labeling process was carried out by two annotators (the first two authors), each independently labeling the entire dataset once. These labels were then merged to form a final label for each app review. In rare cases of disagreement (0.3\% of labels) the annotators reviewed the disputed review together and reached an agreement. Out of 4,762 app reviews, 2,513 were labeled 1 (indicating an accessibility-related concern), and the rest were labeled 0. The final labeled dataset is available on \cite{anonymous_2024_14060778}.

\begin{figure}
    \centering 
    \includegraphics[width=0.5\textwidth]{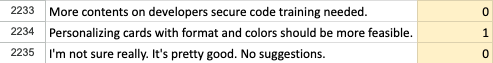}	
    \caption{Three sample app reviews in our dataset.} 
    \label{fig:samplereviews}%
    \vspace{-1em}
\end{figure}

Figure \ref{fig:samplereviews} presents three sample app reviews. The first is labeled 0 because, according to Rule 3, it reflects a developer-side issue. The second is labeled 1 as it refers to the feasibility and accessibility of the app's personalization feature. The third is labeled 0 since the user expresses satisfaction and mentions no issues or inconveniences.

\subsubsection{Dataset preprocessing stages: }
\label{sec:preprocessing}
In the dataset construction phase, we performed several preprocessing methods on the dataset, consisting of correcting spelling errors in app reviews using the TextBlob module\footnote{https://pypi.org/project/textblob/0.9.0/}, eliminating reviews with less than five words, and detecting and removing duplicates.

We then applied data cleaning methods (converting the words to lowercase and removing punctuation and white space), to tokenize and make them more understandable for the models. We utilized Pandas\footnote{https://pandas.pydata.org/} and re\footnote{https://docs.python.org/3/library/re.html} libraries to perform this task.

\subsubsection{Dataset Distribution and Splitting: }

Our final dataset created in Section \ref{sec:dataset} consists of 4,762 app reviews. We divided it using the following stages, using the \textbf{train\_test\_split} module from Python's library, \textbf{sklearn}, by setting the proportion of test size to 10\% of the whole dataset. 

\underline{Training and Validation}: This part incorporates 4,046 app reviews, with 2,166 being accessibility-related reviews and the rest being non-related to accessibility concerns.

\underline{Test}: This part included 716 app reviews, with 347 reviews containing accessibility-related issues and the rest being non-related to accessibility concerns. The app reviews in this part of the dataset were merely used for the final evaluation of the model; therefore, the model did not access these data during the training process.

\begin{figure*}
\setlength\intextsep{6pt}
\setlength\lineskip{0pt}
  \includegraphics[width=\textwidth]{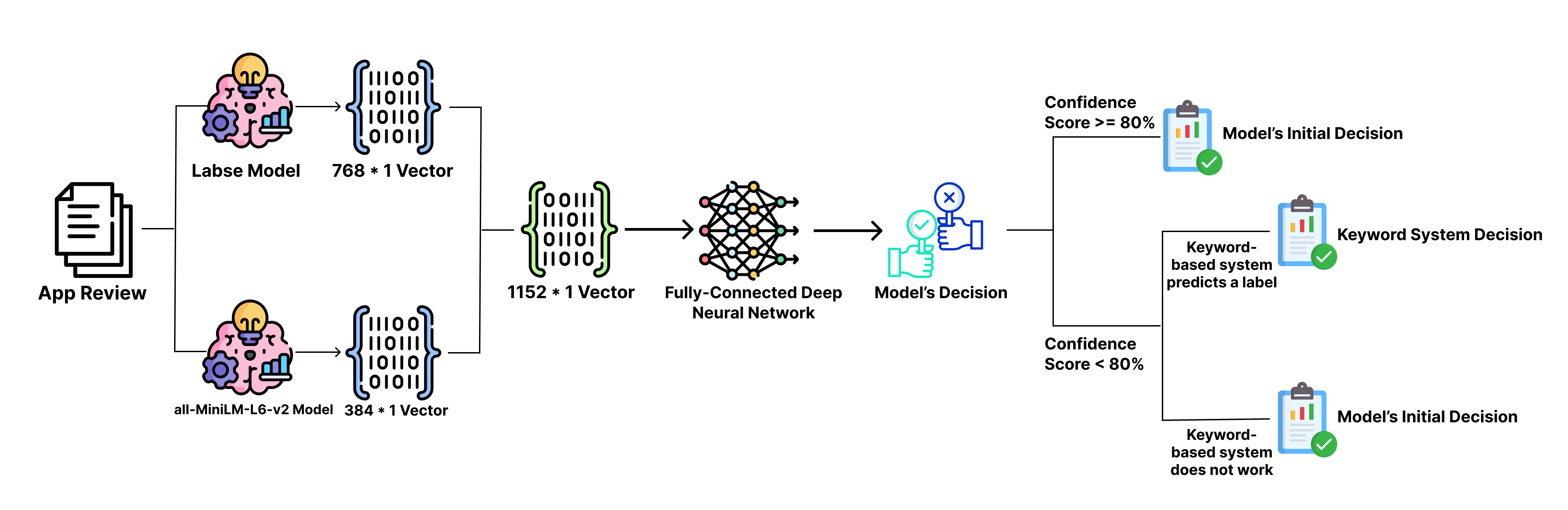}
  \caption{The overall pipeline of our proposed method.}
  \label{fig:overview}
\end{figure*}

\subsection{The Core Architecture of Our Model}

\subsubsection{Overview of the prediction model: }

This section discusses an overview of our designed model for detecting accessibility-related issues in low-code app reviews. The pipeline of our model is shown in Figure \ref{fig:overview}. Each review is first preprocessed by using NLP modules. Then, it is fed to 2 embedding-generator models, leading to the formation of a vector representing the meaning of the review. Consequently, the generated vector goes through a deep neural network for classification, and the network outputs a number, representing the label of the review. Based on our model's confidence score, we leverage our keyword-based system to assist in determining the review's label. In the upcoming sections, each component of our proposed model will be explained in detail.

\subsubsection{Sentence-transformers and Neural Network: }

To generate a meaningful contextual embedding of reviews' text, we used the \textbf{all-MiniLM-L6-v2} model\footnote{https://huggingface.co/sentence-transformers/all-MiniLM-L6-v2} and \textbf{LaBSE} model \cite{feng2020language} from BERT's sentence-transformers module\footnote{https://huggingface.co/sentence-transformers}. These models take sentences as input and then output a 384-dimensional and 768-dimensional vector, respectively. These two vectors are concatenated to form a 1152-dimensional vector, representing the embedding of reviews. Then, the embeddings are fed to a fully connected neural network consisting of 5 layers with activation functions, eventually reducing the count of neurons from 1152 to 2.

\subsubsection{Neural Network Architecture and Training Process: }

After obtaining the app review embeddings, we classified them using a deep, fully connected neural network consisting of five layers: a 1152→512 layer with ReLU activation, a 512→128 layer with PReLU activation, a 128→32 layer with ReLU, a 32→8 layer with ReLU, and a final 8→2 output layer.

For training, we used \textbf{Cross Entropy} as our loss function and \textbf{Adam} as our optimizer. For the model's hyperparameters, we chose the number of epochs as \textbf{3}, the learning rate as \textbf{0.005}, and the batch size as \textbf{32}. Also, we dedicated 10\% of our training/validation dataset to validation and the rest to training. We achieved the best results on the validation data with these hyperparameters, so we chose them as our model's final configuration.

\subsubsection{Keyword-based System: }
Alongside the deep learning model, we implemented a keyword-based system to enhance the model's decision-making process. We provided two sets of keywords: one consisting of keywords representing the concept of accessibility, and another for identifying developer-side issues. These keywords assist the model to label reviews with labels 1 and 0, respectively.

We extracted these keywords by analyzing the reviews containing accessibility issues. We utilized YAKE\footnote{https://pypi.org/project/yake/} to extract common keywords among those reviews, and then we tailored them to represent our custom problem specifically. We created the second set of keywords, indicating software development terminologies, so that we could exclude development-related issues and focus on user-side issues.

Once the model predicts a label for each review, we obtain the confidence score of its decision alongside the predicted label from PyTorch. The confidence score is a value indicating how confident the model is when making a decision. If the confidence score exceeds 80\%, the model's prediction is accepted as the final decision. Otherwise, we use the keyword-based system for reviewing the keywords to check their presence in the text. If a keyword from the first set is found in the review, it is labeled as 1. Conversely, if a keyword from the second set is detected, the review is labeled as 0. If no keyword is found, the model's initial prediction is retained as the final label. This approach seamlessly integrates a traditional keyword-based system with a smart model to analyze reviews effectively. The codes used in this section are available on \cite{anonymous_2024_14060778}.

\section{Evaluation and Results}
In this section, we present the evaluation results of our model for detecting accessibility-related issues in low-code app reviews. We compare our model against three fine-tuned Transformers-based models based on four key metrics which are frequently used for classification tasks: accuracy, recall, precision, and F1-score. The results are displayed in Table \ref{text:accuracies}.

\subsection{Baseline Models}
In this section, we provide more details regarding the models we used as our baselines in the experiment.

\textbf{\textit{BERT}} (Bidirectional Encoder Representations from Transformers) \cite{devlin2018bert} is a seminal model that introduced a bidirectional approach to contextual word representation, using self-attention to better capture word meaning within context.

\textbf{\textit{RoBERTa}} (Robustly optimized BERT approach) \cite{liu2019roberta} builds on BERT’s architecture but modifies the training process to boost performance on NLP tasks, especially with larger datasets and longer sequences. It also uses Masked Language Modeling for pretraining and outperforms BERT in various scenarios, making it a strong benchmark for our evaluation.

\textbf{\textit{DistilBERT}} is a smaller, faster, and more efficient version of BERT \cite{sanh2019distilbert}. Using knowledge distillation, it transfers knowledge from a larger model like BERT into a compact one. This reduces size and computation needs while maintaining much of the original performance.

Additionally, in our pilot studies, we evaluated LLMs (Gemini\cite{team2023gemini} and GPT \cite{achiam2023gpt}) on a subset of 200 app reviews to detect accessibility-related issues. However, the models often produced inconsistent outputs, provided different answers for the same input, and frequently showed uncertainty regarding the presence of accessibility issues. So, we decided to use Transformers models as the backbone of our approach to interpret the reviews and generate their embeddings.

\subsection{Performance Analysis}
Our hybrid model achieves the highest accuracy and F1-score, confirming its effectiveness in detecting accessibility-related issues in app reviews. Notably, it outperforms individual Transformers-based models, especially in recall (82.70\%), showing its strength in correctly identifying accessibility concerns.
Among the baseline Transformers models, \textbf{RoBERTa} performs best, with 70.53\% accuracy and a 74.17\% F1-score. While \textbf{BERT} shows high precision, its lower recall suggests it misses many accessibility-related reviews due to a conservative approach. \textbf{DistilBERT} offers efficient computation and a balanced trade-off but still underperforms compared to our model in all metrics except precision.
Our model’s superior performance is due to its hybrid design, combining Transformers-based models with a keyword-based component. Compared to the Hybrid (No Keywords) model, which relies solely on the Transformers-based predictor, adding the keyword-based component improves metrics (especially recall) by capturing patterns that Transformers might miss. This fusion enables our model to detect accessibility-related issues more effectively than standalone methods.
\vspace{0.4cm}

\begin{table}[!h]
        \vspace{-0.5cm}

    \centering
      \caption{The evaluation metrics on the test set, using different approaches.}
    \small
    \begin{tabular}{c|c|c|c|c}
     \toprule
    \textbf{Classification Model} & \textbf{Accuracy}  & \textbf{Recall} & \textbf{Precision} & \textbf{F1} \\ \hline
    \textbf{Fine-tuned BERT} & 59.21\% & 54.79\% & \textbf{90.48}\% & 68.26\%       \\ \hline
    \textbf{Fine-tuned RoBERTa} & 70.53\% & 64.46\% & 87.31\% & 74.17\%     \\ \hline
    \textbf{Fine-tuned DistilBERT} & 62.04\%  & 62.04\% & 83.86\% & 71.32\%     \\ \hline
    \textbf{Hybrid Model} & \textbf{78.07}\% & \textbf{82.70}\% & 74.73\% & \textbf{78.52}\%     \\ \hline
    \textbf{Hybrid (No Keywords)} & 74.67\%  & 77.82\% & 74.06\% & 75.89\%  \\    \bottomrule
    \end{tabular}
    \label{text:accuracies}
    \vspace{-0.5cm}
\end{table}

\subsection{Error Analysis}
While our hybrid model achieves high overall performance, we observe that its precision is lower than that of the fine-tuned BERT model. This suggests that our model is slightly more prone to false positives, classifying some non-accessibility-related reviews as containing accessibility concerns. 
Conversely, the baseline Transformers models, particularly BERT and RoBERTa, tend to prioritize precision over recall, meaning they are more conservative in labeling reviews as accessibility-related. However, this comes at the cost of missing a significant number of accessibility-related concerns.

\section{Related Work}
\subsection{Software Accessibility}
Accessibility plays a crucial role in software development and mobile applications. Petire et al.\cite{petrie2015towards} define web accessibility as ensuring that all users, particularly those with disabilities or older adults, can access websites across various contexts, including both mainstream and assistive technologies. A survey by Ballantyne et al.\cite{franco2018survey} highlights the lack of automated solutions, posing challenges for developers facing time-to-market constraints. Their findings underscore the need for improved tools to support accessibility evaluations. Similarly, Kavcic~\cite{kavcic2005software} has proposed recommendations and guidelines to point out the difficulties and obstacles that most software applications pose to disabled people and to seek guidelines on accessible software design.

With the growing adoption of mobile devices, Fioravanti et al.~\cite{freire2019accessibility} reviewed accessibility features in mobile learning applications for elderly users, identifying significant gaps and emphasizing the need for the inclusive design as the population ages.

Research has also examined assistive technologies in software development process. A case study by Freire et al.\cite{freire2019evaluation} on a global IT company found that employees with disabilities preferred tools that improved convenience over those offering extensive assistance, highlighting usability and awareness gaps, and suggesting a need for better integration and training for developers during the development process. Mehralian et al.\cite{mehralian2022too} have used assistive technologies to develop an automated framework to dynamically detect and verify overly accessible elements in mobile applications.

In the context of low-code platforms, a systematic literature review by Khalajzadeh et al.~\cite{KHALAJZADEH2025107570} concludes that research on accessibility in low-code environments remains limited. It suggests that future studies should prioritize improving accessibility in low-code environments to better support diverse users.

\subsection{Accessibility Discussions in App Reviews}  
Recent research has tried to identify frustrating and problematic mobile app features using user reviews \cite{wang2022your}. Among these, accessibility-related issues often remain overlooked, negatively affecting end-users. Large-scale studies have identified significant accessibility issues in Android apps. Alshayban et al.~\cite{alshayban2020accessibility} analyzed 1,000 apps and found widespread flaws, revealing that many developers lack accessibility knowledge and that organizations often disregard these concerns. Additionally, app ratings fail to reflect accessibility issues, as users with disabilities constitute a small portion of the user base, highlighting the need for better accessibility education and development tools.  

Several studies have leveraged app reviews to detect accessibility issues. Anam et al.~\cite{anam2013accessibility} analyzed review content to infer accessibility-related concerns and their sentiment. AlOmar et al.~\cite{alomar2021finding} introduced a method for automatically detecting accessibility issues in software application reviews using traditional machine learning and neural networks. Similarly, Tamjeed et al.~\cite{tamjeed2020accessibility} developed a system that identifies accessibility-related concerns by extracting keyword-based features from user reviews. While these approaches significantly contribute to identifying accessibility issues in app reviews, our model employs more advanced techniques, and it specifically targets reviews of low-code applications.

Despite these efforts, accessibility remains a low priority in mobile app development. An analysis of Google Play Store reviews found that only 1.2\% mentioned accessibility, and even when issues were noted, app ratings remained high, suggesting a lack of user and developer prioritization \cite{eler2019android}. Furthermore, research on crowdsourced software engineering task recommendations suggests that integrating human factors into recommendation systems could enhance mobile app accessibility \cite{nirmani2024systematic}. Collectively, these studies highlight a critical gap in accessibility and emphasize the potential of user-driven insights as a solution.

\section{Discussion}
Our study explores identifying accessibility-related issues in low-code app reviews using a hybrid model that combines Transformers-based models with a traditional keyword-based system. Our proposed hybrid model achieved 78\% accuracy and F1-score, demonstrating its effectiveness. Results show that combining keyword-based and Transformers-based approaches yields a more robust solution than using either alone. The keyword system captures explicit accessibility terms, while Transformers-based model better detects implicit issues. This indicates hybrid models are valuable for text classification tasks requiring both linguistic patterns and contextual understanding. Moreover, using LaBSE improves generalization across linguistic variations, enhancing adaptability to diverse feedback.

\subsection{Implications for low-code end-users}
Prior studies on accessibility detection often focus on structured feedback, such as GitHub issues or curated reports \cite{bi2021first, zhao2024accessibility}, while our work explores unstructured, user-generated app reviews. Compared to purely keyword-based methods, our hybrid model captures more nuanced accessibility concerns. For users, especially those with visual impairments, our tool reveals accessibility issues that might otherwise go unnoticed. By automatically identifying and categorizing these issues, the tool enables developers to address barriers, making low-code platforms more inclusive. This can improve usability and allow a broader range of users, including those with visual impairments, to effectively engage with low-code development. Users are also more likely to trust and engage with these tools. A responsive development process that integrates accessibility feedback fosters a community where users feel heard and valued.

\subsection{Implications for low-code practitioners}
The high prevalence of accessibility-related complaints underscores the need for low-code platform developers to prioritize accessibility concerns. Manually filtering through a huge number of app reviews to identify accessibility concerns can be time-consuming and inefficient. Our model provides an automated means to monitor and analyze user feedback, enabling developers to proactively address accessibility issues, and to prioritize accessibility improvements in their design and development workflows. This proactive approach helps developers comply with accessibility standards, reduce the risk of excluding users, and foster a more diverse user base. As accessibility regulations and guidelines (such as WCAG \cite{caldwell2008web}) become more stringent, having a tool that flags accessibility-related issues in user feedback can assist developers in meeting compliance requirements. This can prevent legal risks and enhance the reputation of low-code platforms as being inclusive and user-friendly.

\subsection{Threats to Validity}
In conducting this study, we acknowledge a range of threats to its validity that spans three different validity concerns.

\textbf{\textit{Internal Validity Threats:}} Although we have employed and tested top-performing models from the literature relevant to our work (DistilBERT and RoBERTa), a key limitation of our work is the lack of evaluation with more state-of-the-art models, which could provide a stronger baseline for comparison. Additionally, while our keyword-based system enhances detection, its effectiveness depends on the completeness and quality of the chosen keywords. Expanding and refining this keyword set could improve accuracy and reduce potential biases in detecting accessibility-related issues.

\textbf{\textit{External Validity Threats:}} While our dataset of 4,762 app reviews is well-distributed, a larger dataset could further enhance the model's robustness and generalizability. Future work could address these gaps by collecting more diverse app reviews and testing the approach across different software domains.

\textbf{\textit{Construct Validity Threats:}} Although our hybrid model effectively performs our classification task, reliance on predefined keywords may lead to false positives or missed issues, because our set of keywords is limited and may not include all issues. Our dataset of 4,762 balanced reviews helps mitigate bias, but potential subjectivity in labeling could still affect results. Lastly, our model’s generalizability depends on dataset diversity, and future work could improve robustness by incorporating a broader range of accessibility-related complaints.

\section{Conclusion and Future Directions}
In this work, we proposed a novel intelligent model to detect accessibility-related issues in low-code app reviews. Our model combines state-of-the-art Transformers-based models with a traditional keyword-based system to effectively analyze user feedback. Additionally, we constructed a dataset of 4,762 low-code app reviews by integrating existing datasets with newly collected reviews. Although our model has certain limitations, it serves as a foundational step toward enhancing accessibility awareness and monitoring in low-code platforms.

Several directions can be explored in future research. First, expanding the dataset with a larger and more diverse set of low-code app reviews could improve the model’s robustness and generalizability. Second, qualitative studies involving users of low-code apps with accessibility needs could provide deeper insights into real-world challenges. Third, incorporating sentiment analysis techniques could help distinguish between minor inconveniences and critical accessibility issues. Finally, deploying our model as a real-time monitoring tool for low-code platforms could offer valuable insights for developers, enabling them to proactively address accessibility concerns and improve user experience.

\bibliographystyle{ACM-Reference-Format}
\bibliography{references}

\end{document}